\documentclass[reprint,twocolumn,superscriptaddress,showkeys,
nofootinbib,notitlepage,amsmath,amssymb,floatfix,aps,prd]{revtex4-2}

\usepackage{latexsym,amsmath,amssymb,lmodern,float,url}
\usepackage{natbib}
\usepackage{color}
\usepackage{multirow}
\usepackage{bm}
\usepackage{array}
\usepackage[normalem]{ulem}
\usepackage{cleveref}
\usepackage{xcolor}

\usepackage{tikz}
\usetikzlibrary{datavisualization}
\usetikzlibrary{datavisualization.formats.functions}
\usetikzlibrary{plotmarks}

\usepackage{mathtools}
\def\de{\delta^{\vphantom{1}}}
\def\bde{{\bar\delta}}

\def\h3{{\displaystyle{\frac 3 2}}}

\begin{document}
\title{Hidden-Strangeness Tetraquarks in the Dynamical Diquark Model}
\author{Shahriyar Jafarzade}
\email{sjafarz2@asu.edu}
\author{Richard F. Lebed}
\email{Richard.Lebed@asu.edu}
\affiliation{Department of Physics, Arizona State University, Tempe,
AZ 85287, USA}
\date{May, 2025}

\begin{abstract}
The dynamical diquark model describes multiquark exotic hadrons in terms of diquark components nucleated by heavy quarks, and successfully explains multiple features of hidden-charm and -bottom exotics.  Here we apply the model to the marginally heavy case of hidden-strange states to probe whether mesons near 2~GeV with peculiar properties, such as $\phi(2170)$, $f_2(2340)$, and $X(2370)$, are possible tetraquark candidates.  We calculate spin-multiplet average masses using potentials obtained through lattice simulations and quark models, and we also describe the detailed spectra of the expected multiplets as a diagnostic to discern the nature of future hadrons likely to be discovered in this mass region by experiments at facilities such as BESIII, JLab, and the EIC.
\end{abstract}

\keywords{Exotic hadrons, diquarks}
\maketitle

\section{Introduction}\label{sec:Intro}

One of the greatest advances of particle physics in the 21$^{\rm st}$ century has been the discovery and confirmation of scores of exotic multiquark hadrons~\cite{ParticleDataGroup:2024cfk}.  Most of these states appear in the hidden-charm sector (both nonstrange and strange, both tetraquarks and pentaquarks), but others have been identified in the bottomoniumlike system and even in the open di-charm system ($T_{cc}^+$,~\cite{LHCb:2021vvq,LHCb:2021auc}).

All of these states share the property of containing heavy ($c,b$) quarks, a feature that has been instrumental in identifying them as exotic.  For example, a state like $Z_c(3900)^+$ with mass that lies among the conventional charmonium states (and which decays to conventional charmonium, in this case $J/\psi$) but that also carries nonzero electric charge is clearly exotic, since its minimal valence-quark content is $c\bar c u\bar d$.  Likewise, quark potential models have quite successfully predicted the entire $c\bar c$ spectrum up to the open-flavor $D \bar D$ threshold, and so the numerous $J^{PC} = 1^{--}$ states appearing in addition to the predicted conventional charmonium $\psi$ states [such as $Y(4230)$] are interpreted as exotic.

Part of this success in the heavy-quark sector is by experimental design (the $B$-factory experiments BaBar and Belle produced large, clean samples of $B \to c\bar c$ decays; BESIII is a tau-charm factory; LHCb is constructed to produce enormous data sets of hadronic decays).  But of course, huge samples of lighter ($u,d,s$) hadrons have been collected for decades, and yet precious few clear signals of exotic hadrons have emerged.\footnote{Exceptions include the $J^{PC}$ (=$1^{-+}$)-exotic $\pi_1 (1600)$ and the likely glueball-dominated $f_0(1710)$.}  The quark-flavor universality of QCD demands that any dynamical phenomena observed for heavy quarks should appear for light quarks as well.  But the appearance of heavy quarks $Q$ provides several advantages over their lighter counterparts in the production of experimentally identifiable exotic candidates~\cite{Brambilla:2022ura}: They satisfy $\Lambda_{\rm QCD}/m_Q \ll 1$, which leads to relatively larger mass splittings between multiplets and hence more cleanly separated states (whereas light-quark hadrons form a dense forest of states up to $\approx 2.5$~GeV); their Fermi kinetic energy is suppressed by $\Lambda_{\rm QCD}/m_Q$, making it easier for the heavy quarks to nucleate clusters with the light quarks, as well as often suppressing decay widths; and as noted above in the case of $Z_c(3900)^+$, the simple presence of both heavy and light quarks allows more valence-quark combinations that can be unambiguously identified as exotic.

The utility of heavy quarks in forming identifiable multiquark exotics motivates the development of the {\it dynamical diquark model}~\cite{Brodsky:2014xia,Lebed:2017min}.  Here, the heavy quarks nucleate diquarks $\de \equiv (Qq)_{\bf \bar 3}$ and antidiquarks $\bde \equiv (\bar Q \bar q)_{\bf 3}$ via the attractive ${\bf 3} \! \times \! {\bf 3} \! \to \! \bar {\bf 3}$ short-distance color interaction (and its conjugate).   As long as the $\de$ and $\bde$ quasiparticles form with characteristic sizes smaller than the $\de$-$\bde$ separation---which can be realized, for example, in high-energy collider environments---then the full resulting tetraquark state can assume a $\de$-$\bde$ configuration.  Likewise, pentaquarks can be formed in this model through the formation of a diquark and a {\it triquark}, $\bar \theta \equiv \left[ \bar Q \left( q_1 q_2 \right)_{\bar{\bf 3}} \right]_{\bf 3}$~\cite{Lebed:2015tna}.

This particular implementation of diquark substructure within multiquark hadronic states differs greatly from its original application~\cite{Jaffe:1976ig}, which addressed peculiarities in the spectrum of light scalar mesons such as $f_0(980)$ and $a_0(980)$.  For example, the distinction between ``good'' ($J^P = 0^+$) and ``bad'' ($J^P = 1^+$) light diquarks~\cite{Jaffe:2004ph} central to studies of light tetraquarks is suppressed by effects of $O(1/m_Q)$ for diquarks containing a heavy quark $Q$.

The original dynamical diquark model has successfully been applied to multiple flavor and spin sectors: the $c\bar c q\bar q^\prime$ multiplet average masses~\cite{Giron:2019bcs}, the fine structure of the $c\bar c q\bar q^\prime$ ground-state~\cite{Giron:2019cfc} and orbitally excited~\cite{Giron:2020fvd} multiplets, $b\bar b q \bar q^\prime$ and $c\bar c s\bar s$~\cite{Giron:2020qpb}, $c\bar c c\bar c$~\cite{Giron:2020wpx} and other all-heavy tetraquarks~\cite{Mutuk:2022nkw}, $c\bar c s\bar q$~\cite{Giron:2021sla}, $T_{cc}^{++}$~\cite{Mutuk:2024vzv}, and $c\bar c \, q_1 q_2 q_3$ pentaquarks~\cite{Giron:2021fnl}.  In each case where data are available, the model is supported by experimental findings; furthermore, it provides multiple predictions for yet-unseen states.

However, as already noted, the presence of heavy quarks is central to the argument for the physical plausibility of the dynamical diquark model.  The key question to be addressed in this paper is the extent to which the model can provide a meaningful description of multiquark exotic states (specifically tetraquarks) in which the heaviest quarks are strange.  Of course, since $m_s/\Lambda_{\rm QCD} = O(1)$, the usual arguments are no longer valid, and thus this work provides tests of the most marginal cases to which the original $\de$-$\bde$ picture might apply. Nevertheless, lattice-QCD studies such as~\cite{Dudek:2013yja,Gui:2021vhf} have demonstrated the existence of hybrid mesons with strange-quark content that feature explicit gluonic excitations coupled to the $s\bar s$ pair.  While such simulations do not explicitly invoke a color flux-tube picture, they clearly show that nontrivial gluonic degrees of freedom persist in the strange sector and contribute to hadron structure.  These results support the physical plausibility of extended gluonic fields mediating interactions beyond simple quark models, which is a key assumption underlying Born-Oppenheimer-inspired  frameworks such as the dynamical diquark model.  Although hybrids and tetraquarks differ in their quark content and color structure, the demonstrated persistence of gluonic-excitation effects lends credence to the existence of exotic tetraquark states with strange quarks that utilize this structure.

Such states are expected to have masses in the range of $2.0$--$2.5$~GeV, and are particularly relevant to the experimental programs of BESIII, JLab (particularly GlueX), and eventually the EIC\@.  They likely coexist amongst excited conventional $s\bar s$ mesons, $s\bar s g$ hybrids, and glueballs, making a detailed study to disentangle these possibilities a project of both current theoretical and experimental interest.  Since our numerical inputs ({\it e.g.}, diquark masses) are taken from calculations using other approaches that respect QCD dynamics (lattice, QCD sum rules, quark models), our primary message is: If the flavor universality of QCD allows the same dynamical mechanisms to persist in the strange sector, then this work gives their spectrum and the mass range in which to look for them.

This paper is structured as follows.  In Sec.~\ref{sec:Data} we enumerate and discuss experimentally observed mesons in the region of 2~GeV that potentially could contain a significant $s\bar s s\bar s$ or $s\bar s q\bar q$ component, such as $\phi(2170)$, $f_2(2340)$, and $X(2370)$.  Section~\ref{sec:Model} describes the dynamical diquark model, the formalism in which the $s\bar s s\bar s$ and $s\bar s q\bar q$ systems are analyzed, and introduces the minimal set of fine-structure effects responsible for the level structure of these systems.  The numerical inputs used to produce our phenomenological analysis, as well as the detailed results of that analysis, appear in Sec.~\ref{sec:Results}.  We summarize our findings and describe future prospects for the study of these states in Sec.~\ref{sec:Concl}.

\section{Possible $s\bar s s\bar s$ and $s\bar s q\bar q$ Candidates}\label{sec:Data}

The (unflavored) hadron spectrum above 2~GeV presented in the Particle Data Group (PDG) Summary Table~\cite{ParticleDataGroup:2024cfk} contains several meson resonances that challenge the traditional quark-antiquark ($q\bar{q}$) picture.  Isoscalar meson resonances such as $\phi(2170)$ (hidden $s\bar s$, $J^{PC} = 1^{--}$), $f_0$ ($0^{++}$), and $f_2$ ($2^{++}$) may incorporate exotic structures such as glueballs, hybrids, or tetraquarks. 

The state $\phi(2170)$ [or $Y(2175)$] was first observed by the BaBar Collaboration~\cite{BaBar:2006gsq} in the process $e^{+}e^{-}\rightarrow \phi f_0(980)$, and was subsequently confirmed by BES~\cite{BES:2007sqy} and Belle~\cite{Belle:2008kuo}.  This resonance has been confirmed by the BESIII Collaboration in several independent channels, notably $e^+e^- \to \phi f_0(980)$~\cite{BESIII:2021aet}, $\phi \, \eta$~\cite{BESIII:2021bjn}, and $\phi \, \eta'$~\cite{BESIII:2020gnc}.  These channels exhibit clear resonant structures around 2.17~GeV in the invariant mass spectra, in agreement with the earlier observations by BaBar, BES, and Belle.  The consistency of mass and width measurements across these processes,
%---typically around 2.17~GeV and 80--100~MeV,
summarized by the PDG as $2146 \pm 6$~MeV and $106^{+24}_{-18}$~MeV, respectively~\cite{ParticleDataGroup:2024cfk}, strongly supports the existence of this resonance.  The BESIII results suggest a nontrivial internal structure for $\phi(2170)$, making it a strong candidate for exotic interpretations such as a tetraquark state or a hybrid meson, rather than as a simple radial excitation of the vector meson $\phi(1020)$.  Its internal structure has been shown to conflict with a hybrid-meson hypothesis in the analysis of Ref.~\cite{BESIII:2020gnc}.  In addition, the mass of the lowest $1^{--}$ glueball candidate in lattice QCD (LQCD) is estimated to be much higher, around 3.8~GeV~\cite{Morningstar:1999rf,Chen:2005mg,Athenodorou:2020ani}.

The \(2^{++}\) \( f_2(1910) \) and \( f_2(1950) \) resonances have long been considered potential non-\( q\bar{q} \) candidates.  $f_2(1950)$ in particular is a well-established tensor resonance with an unusually large width of \(464 \pm 24\)~MeV\@.

The observations that the resonance $f_0(2020)$ decays to $\eta^\prime\eta^\prime$~\cite{BESIII:2022zel}, and that $f_2(2010)$ decays to $\bar{K}K$~\cite{Vladimirsky:2006ky} and $\phi \, \phi$~\cite{BESIII:2016qzq}, suggest that both states likely possess significant $s\bar s$ content.  The proximity of the $\phi \, \phi$ threshold at 2.039~GeV indicates the possibility that one or both of the resonances could be either a $\phi \, \phi$ molecular state or a hybrid meson resulting from the gluonic excitation of an $s\bar{s}$ pair.

The tensor ($J^{PC} = 2^{++}$) resonances $f_2(2300)$ and $f_2(2340)$  are of distinct interest in the study of exotic hadrons, especially glueballs.  They have both been observed in radiative $J/\psi$ decays, particularly in channels such as $J/\psi \to \gamma \phi\phi$~\cite{BESIII:2016qzq} and $\gamma \eta\eta$~\cite{BESIII:2013qqz}, where substantial gluonic couplings are expected.  In particular, LQCD simulations~\cite{Morningstar:1999rf,Chen:2005mg,Athenodorou:2020ani} predict the lightest tensor glueball to lie near 2.4~GeV, and the masses of these $f_2$ states are consistent with such expectations.  On the other hand, the observed decay channels of $f_2(2300)$, namely $\phi\, \phi$, $\Lambda\bar{\Lambda}$, and $K\bar{K}$, suggest a significant $s\bar s$ content.  The resonance $f_2(2340)$ exhibits a notably broad width (331~MeV), suggesting an exotic nature beyond the conventional $q\bar q$ picture.  Nevertheless, the measured production rate of $f_2(2340)$ radiative decays such as via $J/\psi \to \gamma \eta^\prime \eta^\prime$~\cite{BESIII:2022zel} is significantly lower than the production rate expected from LQCD for a pure tensor glueball~\cite{Yang:2013xba}.  
Recent phenomenological analysis~\cite{Vereijken:2023jor} based upon a chiral effective model~\cite{Giacosa:2024epf} suggests that another broad $2^{++}$ state, $f_2(1950)$, could be a glueball-rich resonance.  Despite such insights, a definitive interpretation of these states remains elusive due to an abundance of overlapping resonances combined with limited experimental resolution.

We also comment on the $X$ states with masses $\leq 2.5$~GeV observed by the BESIII Collaboration.  The state $X(2370)$ is given in the PDG particle listing, but not in the Summary Table.  Initial evidence for this state emerged in 2011 through analysis of the decay channel $J/\psi \to \gamma \pi^+\pi^-\eta'$~\cite{BESIII:2010gmv}.  Subsequent studies by BESIII have confirmed the existence of $X(2370)$ in additional decay modes, including $J/\psi \to \gamma K^+ K^- \eta'$ and $J/\psi \to \gamma K_S^0 \bar K_S^0 \eta'$~\cite{BESIII:2019wkp}.  These observations allowed for detailed measurements of the resonance's properties such as its mass and width, and determined its $J^{PC}$ to be $0^{-+}$~\cite{BESIII:2023wfi}.  Its mass makes it a plausible candidate for the $0^{-+}$ glueball predicted by LQCD, which occurs near the somewhat higher value of 2.6~GeV~\cite{Morningstar:1999rf,Chen:2005mg,Athenodorou:2020ani}.  The BESIII Collaboration also reported two other resonances with masses closer to the LQCD prediction (with undetermined $J^{PC}$, but $0^{-+}$ being at least somewhat favored): $X(2500)$~\cite{BESIII:2016qzq} and $X(2600)$~\cite{BESIII:2022sfx}.  One further pseudoscalar state in the PDG light-unflavored meson list is \( X(1835) \).  It was first reported by BESII in the radiative decay \( J/\psi \to \gamma p \bar p \)~\cite{BES:2003aic}, later confirmed in \( J/\psi \to \gamma \pi^+ \pi^- \eta' \)~\cite{BES:2005ega}, and its quantum numbers $0^{-+}$ were determined in Ref.~\cite{BESIII:2011aa}.

Continued investigation into hidden-strange tetra\-quark candidates over a broader range of energies and decay modes—--beyond just the $\phi \, \pi$ channel~\cite{BESIII:2018rdg} in the case of isovector ($Z_s$) candidates—--is crucial for disentangling their internal structure, and for distinguishing them from conventional or hybrid $s\bar s$ resonances.

\section{The Dynamical Diquark Model}\label{sec:Model}

The \emph{dynamical diquark model}~\cite{Brodsky:2014xia,Lebed:2017min} provides a theoretical framework for describing exotic hadrons---such as tetraquarks and pentaquarks---as bound systems of compact diquarks and antidiquarks (or triquarks), dynamically confined by the gluonic field.  Unlike static constituent models, this approach treats the relative motion of the color sources explicitly, allowing a more realistic description of spatial separation and interaction dynamics.

A central ingredient of the model is the application of the \emph{Born-Oppenheimer} (BO) \emph{approximation}, in which the quickly changing gluonic field is treated separately from the slower motion of the heavy constituents.  For a given gluonic excitation labeled by quantum numbers $\Lambda_\eta^\epsilon$ ({\it e.g.}, $\Sigma_g^+$, $\Pi_u^-$)~defined later in this section, the diquark–antidiquark pair at separation $r$ moves in an effective potential $V_{\Lambda_\eta^\epsilon}(r)$ determined by LQCD simulations.

To calculate the mass spectrum of such exotic states, one solves the radial Schrödinger equation for the relative motion of the diquark ($\de$) and antidiquark ($\bde$) in the corresponding BO potential:
\begin{equation}
\left[ -\frac{1}{2\mu} \frac{d^2}{dr^2} + \frac{L(L+1)}{2\mu r^2} + V_{\Lambda_\eta^\epsilon}(r) \right] R_{nL}(r) = E_{nL} R_{nL}(r),
\label{eq:schrodinger}
\end{equation}
where $\mu = \frac{m_\de m_\bde}{m_\de + m_\bde}$ is the reduced mass of the $\de$-$\bde$ system, $L$ is their relative orbital angular momentum, and $V_{\Lambda_\eta^\epsilon}(r)$ is the BO potential extracted from LQCD for the given gluonic excitation.  The choice of BO potential depends upon the quantum numbers of the gluonic field, which are classified using the symmetries of the cylindrical group $D_{\infty h}$, analogous to those used in diatomic molecular physics.

The total mass of the tetraquark state is then given by:
\begin{equation}
M_0(nL) = m_\de + m_\bde + E_{nL},
\label{eq:mass}
\end{equation}
where $m_\de$ and $m_\bde$ are the constituent diquark and antidiquark masses, and $E_{nL}$ is the eigenvalue obtained from solving Eq.~\eqref{eq:schrodinger}.

In addition to this leading-order (muliplet-average) mass spectrum obtained~\cite{Giron:2019bcs} by solving the radial Schrö\-dinger equation Eq.~\eqref{eq:schrodinger}, fine-structure effects arising from spin-dependent interactions between $\de$ and $\bde$ must be included in order to reproduce the observed multiplet splittings~\cite{Giron:2019cfc,Giron:2020fvd}.  These effects, including spin-spin, spin-orbit, and tensor couplings, are treated perturbatively and induce the following terms within the Hamiltonian:
\begin{equation}\label{eq:hamiltonian}
  H=H_0+2\kappa_{sq}({\bf s}_q \cdot {\bf s}_{s^{\vphantom\dagger}} \! +{\bf s}_{\bar{q}} \cdot {\bf s}_{\bar{s}^{\vphantom\dagger}})+V_{LS} \, {\bf L}\cdot {\bf S}+V_T S^{\delta\bar{\delta}}_{12} \,,  
\end{equation}
with multiplet-average Hamiltonian $H_0$, diquark-spin operators ${\bf s}_\de \equiv {\bf s}_q + {\bf s}_s$, ${\bf s}_\bde \equiv {\bf s}_{\bar q} + {\bf s}_{\bar s}$, total quark spin operator ${\bf S} \equiv {\bf s}_\de + {\bf s}_\bde$, and orbital angular momentum operator ${\bf L}$.  The tensor operator $S_{12}^{\de\bde}$ is defined by
\begin{equation}
\label{eq:Tensor}
S_{12}^{\de\bde} \equiv 4 \left[ 3 \, {\bm s}_\de \! \cdot {\bm r} \, {\bm s}_\bde \! \cdot {\bm r} / r^2 - {\bm s}_\de \! \cdot {\bm s}_\bde \right] \, .
\end{equation}
For simplicity, isospin effects are neglected in this work (and indeed, are irrelevant for $s\bar s s\bar s$ states).

Such operators are well known, and their matrix elements have been computed in many places.  In the context of the dynamical diquark model, they are collected in Refs.~\cite{Giron:2020fvd,Giron:2020wpx}.  For completeness, we restate them here:
\begin{equation}
\Delta M_{\rm spin-spin} = \kappa_{sq} \left[ s_\de (s_\de + 1) + s_{\bde} (s_{\bde} + 1) - 3 \right] \, ,
\end{equation}
\begin{equation}
\Delta M_{LS} = \frac{V_{LS}}{2} [J(J+1)-L(L+1)-S(S+1)] \, ,
\end{equation}
where $J$ is the quantum number corresponding to \( {\bf J} = {\bf L} + {\bf S} \), and
\begin{eqnarray}
\label{eq:S12general}
\lefteqn{\left< L',S',J
\right| S_{12} \left| L,S,J \right>} \nonumber \\
& = & (-1)^{S+J} \!
\sqrt{30[L][L'][S][S']} \left\{ \begin{array}{ccc} J & S' & L' \\ 2 &
L & S \end{array} \right\} \! \left( \! \begin{array}{ccc} L' & 2 & L
\\ 0 & 0 & 0 \end{array} \! \right) \nonumber \\
& & \times \left\{ \! \begin{array}{ccc}
s_\de & s_\bde & S \\ s_{{\de}^\prime} & s_{{\bde}^\prime} & S' \\ 1
& 1 & 2 \end{array} \! \right\} \! \left< s_{{\de}^\prime} ||
\bm{\sigma}_1 || s_{\de} \right> \left< s_{{\bde}^\prime} ||
\bm{\sigma}_2 || s_\bde \right> ,
\end{eqnarray}
where $[j] \! \equiv \! 2j \! + \! 1$, parentheses containing 6 elements denote a $3j$ symbol, and braces denote a $6j$ or $9j$ symbol, as indicated by the number of elements.  The reduced matrix elements of the canonically normalized angular-momentum generators {\bf j} are given by
\begin{equation}
\label{eq:Jreduced}
\left< j^\prime || \, {\bf j} \, || \, j \right> =
\sqrt{j(2j+1)(j+1)} \, \delta_{j^\prime j} \, . 
\end{equation}
The operators $\bm{\sigma}$ in Eq.~(\ref{eq:S12general}) are twice the relevant canonically normalized angular-momentum generators, thus generalizing the familiar relation for spin-$\frac 1 2$ Pauli matrices.

In the dynamical diquark model, each state is characterized by radial quantum number \( n \), orbital angular momentum \( L \), total quark spin \( S \), and total angular momentum \( J \) quantum numbers, where \( {\bf J} = {\bf L} + {\bf S} \).  These quantum numbers are combined with the Born-Oppenheimer (BO) quantum numbers \( \Lambda_\eta^\epsilon \), where \( \Lambda \) denotes the projection of the gluonic angular momentum onto the $\bde$-$\de$ axis {\bf r} (\( \Lambda = 0, 1, 2, \dots \), corresponding to \( \Sigma, \Pi, \Delta, \dots \)), \( \eta = g(u) \) labels even(odd) behavior under $CP$ inversion through the midpoint of {\bf r} (the gluon-field flux tube), and \( \epsilon = \pm \) for \( \Lambda \geq 1 \) is the eigenvalue of reflection through any plane containing {\bf r}.

The full state is expressed in the following form:
\begin{equation}
\left| n\, {}^{2S+1}L_J;\, \Lambda_\eta^\epsilon \right\rangle,
\end{equation}
where the spectroscopic notation \( {}^{2S+1}L_J \) identifies the spin and orbital structure, and \( \Lambda_\eta^\epsilon \) specifies the BO potential.\footnote{Specifically, $L$ now includes not only the relative orbital angular momentum of the heavy constituents, but also the angular momentum of the light degrees of freedom, whose projection along the ${\bf r}$ axis is just $\Lambda$.  The complete treatment of all relevant quantum numbers is presented in Ref.~\cite{Lebed:2017min}.}  Phenomenologically, all known heavy exotics to date can be accommodated just within the lowest orbitals of the ground-state BO multiplet $\Sigma^+_g$, specifically $\Sigma^+_g(1S,1P,2S,2P)$~\cite{Giron:2019bcs}.  Thus, the excited BO potentials ($\Pi^+_g$, {\it etc.}) are relevant to a full analysis of all possible states~\cite{Lebed:2017min} (and represent the hybrid analogues for tetraquark states), but for this work we restrict to only the lowest BO potential, $\Sigma^+_g$.

Then each such state can be labeled by $J^{PC}$, where the parity ($P$) and charge-conjugation ($C$) quantum numbers (for neutral tetraquarks) are given by:
\begin{equation} \label{eq:PC}
P = (-1)^{L}, \qquad C = (-1)^{L + s_{q\bar{q}}+ s_{s\bar{s}^{\vphantom\dagger}}}.
\end{equation}
In the $S$-wave case ($L=0$), all 6 $s\bar s q\bar q$ states (including two states apiece for $J^{PC} = 0^{++}$ and $1^{+-}$) have $P = +1$.  They are denoted in the basis $\left| s_\de , s_\bde \right>_J$ as follows~\cite{Maiani:2014aja}:
\begin{eqnarray}
J^{PC} = 0^{++}: & \ & X_0 = \left| 0_\de , 0_\bde \right>_0 \, , \ \
X_0^\prime = \left| 1_\de , 1_\bde \right>_0 \, , \nonumber \\
J^{PC} = 1^{++}: & \ & X_1 = \frac{1}{\sqrt 2} \left( \left| 1_\de ,
0_\bde \right>_1 \! + \left| 0_\de , 1_\bde \right>_1 \right) \, ,
\nonumber \\
J^{PC} = 1^{+-}: & \ & Z \  = \frac{1}{\sqrt 2} \left( \left| 1_\de ,
0_\bde \right>_1 \! - \left| 0_\de , 1_\bde \right>_1 \right) \, ,
\nonumber \\
& \ & Z^\prime \, = \left| 1_\de , 1_\bde \right>_1 \, ,
\nonumber \\
J^{PC} = 2^{++}: & \ & X_2 = \left| 1_\de , 1_\bde \right>_2 \,. 
\label{eq:Swavediquark}
\end{eqnarray}
The $s_{q\bar q} + s_{s\bar s}^{\vphantom\dagger}$ eigenvalues required by Eq.~(\ref{eq:PC}) to compute the $C$ eigenvalues given in Eqs.~(\ref{eq:Swavediquark}) are obtained by recoupling the spins of the quarks within $s_\de$, $s_\bde$ to the basis $s_{q\bar q}, s_{s\bar s}$ through the application of $9j$ symbols~\cite{Lebed:2017min}:
\begin{eqnarray}
\lefteqn{\hspace{-1em} \left< (s_q \, s_{\bar q}) s_{q\bar q} , (s_s \, s_{\bar s}) s_{\bar s s}
, S \, \right| \left. (s_q \, s_s) s_\de , (s_{\bar q} \, s_{\bar s})
s_\bde , S \right>} & & \nonumber \\
& = & \left( [s_{q\bar q}] [s_{s\bar s}] [s_\de] [s_\bde] \right)^{1/2}
\left\{ \begin{array}{ccc} s_q & s_{\bar q} & s_{q\bar q} \\
s_s & s_{\bar s} & s_{s\bar s} \\ s_\de & s_\bde & S \end{array} \! \right\}
\, . \ \ \label{eq:9jTetra}
\end{eqnarray}

The color-{\bf 3} structure of each diquark, as assumed by the dynamical diquark model, contributes to antisymmetry under particle exchange.  Thus, whenever a diquark contains two identical quarks, such as for the fully strange tetraquark $\de\bde = (ss)(\bar{s}\bar{s})$, the remaining quantum numbers for each of $\de$ and $\bde$ must be completely symmetric under quark exchange in order to satisfy the Pauli exclusion principle.  Assuming that $\de$ and $\bde$ both lie in the spatial ground-state (hence symmetric) configuration, then both are restricted to be spin-symmetric axial-vector diquarks (\(s_\de \! = s_\bde \! = \! 1\)).  In this case, the only surviving 3 $S$-wave states in Eqs.~(\ref{eq:Swavediquark}) for the fully strange tetraquarks are $X_0^\prime, Z^\prime, X_2$.  Note in particular the absence of an $S$-wave $1^{++}$ state.

For the $P$-wave states, \(P = -1\).  The possible total angular momenta are \(J \! = \! |L - S|, ..., L + S\), where the total quark spin $S$ here is just $J$ from the prior $S$-wave case.  For \(S \! = \! 0\) (states based upon $X_0^\prime$), \(J\) can only be 1, and hence one obtains a state [using Eqs.~(\ref{eq:PC})] with $J^{PC} \! = \! 1^{--}$.  For \(S \! = \! 1\) (states based upon $Z^\prime)$, the allowed values are \(J \! = \! 0, 1, 2\), giving states with \(J^{PC} = (0,1,2)^{-+}\).  For \(S \! = \! 2\) (states based upon $X_2$), the allowed values are \(J \! = \! 1, 2, 3\), giving the states \(J^{PC} \! = \! (1,2,3)^{--}\).  We note that two distinct \(1^{--}\) states arise in this multiplet, one from \(S \! = \! 0\) and one from \(S \! = \! 2\).  One thus obtains the following 7 distinct $P$-wave states:
\begin{equation}
J^{PC} = (0,1,2)^{-+},\ (2\times 1,2,3)^{--}  .
\end{equation}
This enumeration of states is identical to that for the $c\bar c c\bar c$ system obtained in Ref.~\cite{Giron:2020wpx}.  In the case of isoscalar $s\bar{s}q\bar{q}$ tetraquarks, the number of possible $P$-wave states increases to 14~\cite{Lebed:2017min}.  This set consists of states with multiple copies: $2\times (0,1,2)^{-+}$, $(2\times 2,4\times 1)^{--}$, in addition to $(0,3)^{--}$, all of which arise from combining distinct total quark-spin eigenvalues $S = 0, 1, 2$ with orbital angular momentum $L \! = \! 1$.  Notably, the exotic quantum numbers $0^{--}$, which are forbidden for both conventional $q\bar q$ and $s\bar s s\bar s$ mesons, become accessible for $s\bar s q\bar q$.

For the $D$-wave states, \(P \! = \! +1\).  For \(S \! = \! 0\) (states based upon $X_0^\prime$), \(J\) can only be 2, and hence one obtains a state [using Eqs.~(\ref{eq:PC})] with $J^{PC} \! = \! 2^{++}$.  For \(S \! = 1\) (states based upon $Z^\prime)$, the allowed values are \(J \! = \! 1, 2, 3\), giving the states \(J^{PC} \! = \! (1,2,3)^{+-}\).  For \(S \! = \! 2\) (states based upon $X_2$), the allowed values are \(J \! = \! 0, 1, 2, 3, 4\), giving the states \(J^{PC} \! = \! (0,1,2,3,4)^{++}\).  We note that two distinct \(2^{++}\) states arise in this multiplet, one from \(S \! = \! 0\) and one from \(S \! = \! 2\).  One thus obtains the following 9 distinct $D$-wave states:
\begin{equation} \label{eq:Dwavestates}
J^{PC} = (0,1,2\times 2,3,4)^{++}, \, (1,2,3)^{+-} .
\end{equation}
This enumeration of states is identical to that for the $c\bar c c\bar c$ system obtained in Ref.~\cite{Giron:2020wpx}.  For the isoscalar \( s \bar{s} q\bar{q} \) system, 7 additional states are anticipated~\cite{Lebed:2017min}: 
\begin{equation}
J^{PC} = ( 1, 2 \times 2, 3 )^{++} , \, ( 1, 2, 3 )^{+-} ,  
\end{equation}
which duplicate $J^{PC}$ values already present in Eq.~(\ref{eq:Dwavestates}).

Further refinements, such as identifying hierarchies in the sizes of fine-structure interactions, serve to improve the predictions, at least in terms of correctly ordering the masses of these states within multiplets of the dynamical diquark model.

\section{Results}\label{sec:Results}

To produce tetraquarks within the dynamical diquark model, we adopt a variant of the Cornell potential that incorporates a Coulombic term at small $\de$-$\bde$ separation $r$, a linear confining term at large $r$, and an offset $V_0$:
\begin{equation}\label{eq: potential}
   V_{\Sigma_{g}^{+}}(r) =V_0-\frac{\alpha}{r}+\sigma r\,,
\end{equation}
where the parameters extracted from LQCD calculations~\cite{Morningstar:2019} (similar results can be obtained from Ref.~\cite{Capitani:2018rox}) and from the Relativistic Quark Model (RQM)~\cite{Ebert:2005xj} are presented in Table~\ref{tab:parameters}.
\begin{table}[ht]
    \centering
    \renewcommand{\arraystretch}{1.5}
    \begin{tabular}{|c|c|c|c|}\hline
      Potential   &  $V_0$ ($\text{GeV}$) & $\alpha$  & $\sigma$ ($\text{GeV}^2$) \\\hline 
    % LQCD~\cite{Capitani:2018rox} & -0.322 & 0.263 &0.214 \\\hline
    LQCD~\cite{Morningstar:2019} & $-0.383$ & 0.297 & 0.216 \\\hline
       RQM~\cite{Ebert:2005xj} & $-0.300$ & 0.948, 0.975  & 0.180\\\hline
    \end{tabular}
    \caption{Numerical values of the parameters in Eq.~\eqref{eq: potential}.}
    \label{tab:parameters}
\end{table}

In accordance with the RQM framework~\cite{Ebert:2005xj}, the value of $\alpha$ depends upon the number of active quark flavors in the $\beta$ function and the specific values of quark masses ($m_q=0.33$ GeV, $m_s=0.50$ GeV); from Ref.~\cite{Ebert:2005xj}, we use $\alpha=0.948$ for the $(sq)(\bar s \bar q)$ diquark potential and $\alpha=0.975$ for the $(ss)(\bar s \bar s)$ diquark potential.  The $(sq)$ diquark mass is approximated using the spin-averaged form:
\begin{equation}
    m^{\rm RQM}_{\delta=(sq)}=\frac{3}{4} m^{S=1}_{\delta = (sq)}+\frac{1}{4} m^{S=0}_{\delta = (sq)} = 1.039 \ {\rm GeV} \, , \label{eq:diquark-mass}
\end{equation}
as obtained from the corresponding diquark masses~\cite{Ebert:2005xj} $m^{S=0}_{\delta = (sq)}=0.948$ GeV and $m^{S=1}_{\delta = (sq)}=1.069$ GeV\@.  Similar values also arise from QCD sum rules~\cite{Wang:2011ab}.

We use the findings of Ref.~\cite{Babich:2007ah} to estimate diquark masses within the context of LQCD\@.  In summary,
\begin{enumerate}
    \item The mass difference between a $0^+$ diquark and pair of quarks $q$ in the chiral limit (with inverse lattice spacing $a^{-1} = 2.12$~GeV) is given by
    \begin{equation} \label{eq:LQCDa}
        a \cdot (m_{\delta}^{S=0} - 2m_q) = -0.10\,.
    \end{equation}
    Approximating the ($sq$) constituent quark mass as $m_q = 0.4$~GeV (roughly averaging $m_q$ and $m_s$), Eq.~(\ref{eq:LQCDa}) yields an estimate for the $0^+ \ (sq)$ diquark mass:
    \begin{equation}
        m^{S=0}_{\delta = (sq)} = 0.588~\text{GeV}\,.
    \end{equation}
    \item The mass difference between the $1^+$ (``bad'') and $0^+$ (``good'') ($sq$) diquarks is given by
    \begin{equation} \label{eq:GoodBadDiff}
        m_{\delta =(sq)}^{S=1} - m_{\delta =(sq)}^{S=0} = 0.162~\text{GeV}\,.
    \end{equation}
Using these values in Eq.~\eqref{eq:diquark-mass}, we obtain
\begin{equation} \label{eq:msqLQCD}
m_{\delta=(sq)}^{\text{LQCD}}=0.710 \ {\rm GeV} \, .
\end{equation}
\end{enumerate}
  
For the $(ss)$ case, we use $m_s = 0.50$~GeV, and estimate the $1^+ \ (ss)$ diquark mass using the given $a^{-1}$ value and Eqs.~(\ref{eq:LQCDa}), (\ref{eq:GoodBadDiff}) [the latter assumed to be applicable also for $(ss)$, since LQCD can treat the two $s$ quarks as distinguishable]:
    \begin{equation}
       m_{\delta=(ss)}^{\text{LQCD}}= m_{\delta = (ss)}^{S=1} = 0.950~\text{GeV}\,.
    \end{equation}
The corresponding RQM diquark mass~\cite{Ebert:2005xj} is
\begin{equation} \label{eq:mssRQM}
m_{\delta=(ss)}^{\text{RQM}}=1.203~{\rm GeV} \, .
\end{equation}

\begin{figure*}[]
    \centering
     \includegraphics[scale=0.8]{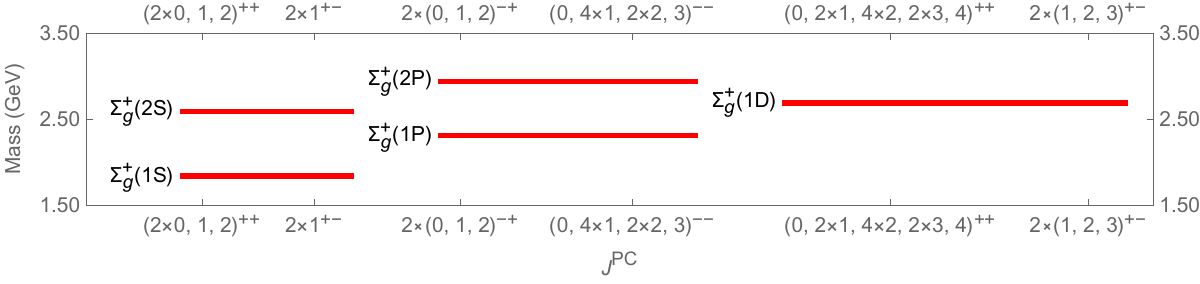}\caption{Spectrum of $s\bar s q\bar q$ tetraquarks within the dynamical diquark model, not yet including fine structure.  The specific mass eigenvalues presented in this figure are computed using the potential of Ref.~\cite{Morningstar:2019}.}
    \label{fig:sq}
\end{figure*}

\begin{figure*}[]
    \centering
    \includegraphics[scale=0.8]{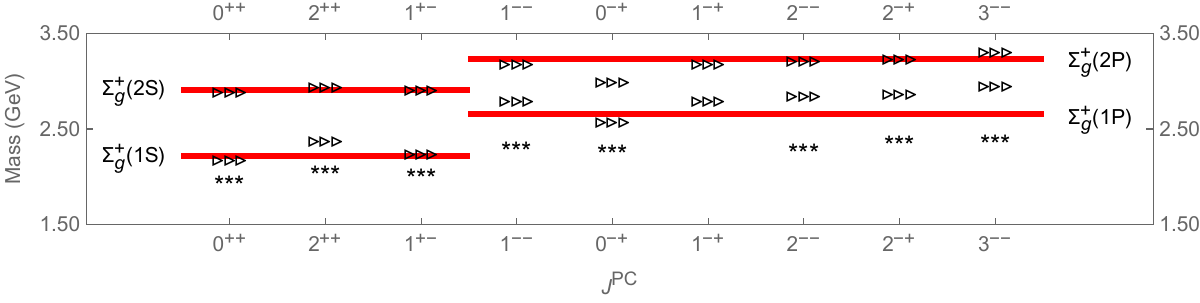}
    \caption{Spectrum of $s\bar s s\bar s$ tetraquarks within the dynamical diquark model (red lines), not yet including fine-structure effects, compared to the results of calculations in a potential model~\cite{Lodha:2024qby} (triangles) and using QCD sum rules~\cite{Su:2022eun} (stars).  The specific mass eigenvalues presented in this figure are computed using the potential of Ref.~\cite{Morningstar:2019}.}
    \label{fig:ss}
\end{figure*}

We compute the tetraquark spectrum within the framework of the dynamical diquark model, using the values in Eqs.~(\ref{eq:diquark-mass}), (\ref{eq:msqLQCD})--(\ref{eq:mssRQM}) of $m_\de = m_\bde$ and the binding energies $E_{nL}$ as inputs to Eq.~\eqref{eq:mass}.  In turn, the $E_{nL}$ eigenvalues are obtained by solving the Schrödinger equation Eq.~\eqref{eq:schrodinger} with the potential $V(r)$ defined in Eq.~\eqref{eq: potential}, using the numerical values for its parameters listed in Table~\ref{tab:parameters}\@.  The resulting spectra for the $(s q)(\bar{q} \bar{s})$ and $(s s)(\bar{s} \bar{s})$ systems (subsequently labeled as $s\bar s q\bar q$ and $s\bar s s\bar s$, respectively) are presented in Tables~\ref{tab:sq-tetraquarks} and \ref{tab:ss-tetraquarks} (and graphically represented in Figs.~\ref{fig:sq} and \ref{fig:ss}), respectively.  These results do not include the spin-dependent fine-structure interactions described by Eq.~\eqref{eq:hamiltonian}, as the corresponding couplings $\kappa_{sq}$, $V_{LS}$, and $V_T$ are currently unknown.  However, to estimate the impact of such effects, we report the parametric mass splittings of $P$- and $D$-wave $s\bar s s\bar s$ tetraquarks in Table~\ref{tab:mass-splittings} [The $S$-wave $s\bar s s\bar s$ states are degenerate, with common mass $M_0(1S) + \kappa_{ss}$], and discuss their pattern later in this section.

\begin{table*}[]
\centering    
\setlength{\extrarowheight}{1.5ex}
\begin{tabular}{cccccccc}
\hline\hline
BO state & Potential & $m_{\delta}$ (GeV) &$M_0$ (GeV)   & $\langle 1/r\rangle^{-1}$ (fm) & $\langle r \rangle$ (fm) &  $\text{Multiplicity}\times J^{PC}$  \\\hline
\multirow{2}{*}{$\Sigma_g^{+}(1S)$} 
& LQCD~\cite{Morningstar:2019} & 0.710  &  1.842 & 0.404  & 0.535  &    \\
 &  RQM~\cite{Ebert:2005xj} & 1.039 & 1.954 & 0.272 & 0.380   & $2 \times 0^{++}, (1,2)^{++}$, $2\times 1^{+-}$ 
   \\\cline{1-6}
\multirow{2}{*}{$\Sigma_g^{+}(2S)$} 
 & LQCD~\cite{Morningstar:2019} & 0.710 &  2.592 & 0.608  & 0.977 &
\\

&  RQM~\cite{Ebert:2005xj} & 1.039 & 2.721 & 0.504 & 0.826  &   
   \\\hline
   
 \multirow{2}{*}{$\Sigma_g^{+}(1P)$} 
 & LQCD~\cite{Morningstar:2019} & 0.710  &  2.311 & 0.696  & 0.802  &    \\
 
 &  RQM~\cite{Ebert:2005xj} & 1.039 & 2.536 & 0.576 & 0.673   & $4\times 1^{--}$,  $2\times 2^{--}$, $2\times (0,1,2)^{-+}$, $(0,3)^{--}$ \\\cline{1-6}
\multirow{2}{*}{$\Sigma_g^{+}(2P)$}

 & LQCD~\cite{Morningstar:2019} & 0.710  &  2.942 & 0.878  & 1.182  &      \\

 &  RQM~\cite{Ebert:2005xj} & 1.039 & 3.082 &0.764 & 1.043 & \\ 
 \hline%\hline
\multirow{2}{*}{$\Sigma_g^{+}(1D)$} 

 & LQCD~\cite{Morningstar:2019} & 0.710  &  2.689 & 0.933 & 1.026 & $4\times 2^{++}$, $2\times (1,3)^{++}$, $(0,4)^{++}$, $2\times (1,2,3)^{+-}$ \\%\cline{2-6} 
 &  RQM~\cite{Ebert:2005xj} & 1.039 & 2.895 & 0.815 & 0.901  & \\\hline\hline
\end{tabular}
 \caption{Masses $M_0$ and corresponding $J^{PC}$ eigenvalues of $s\bar s q\bar q$ dynamical-diquark states, computed using the given diquark masses $m_{\delta}$ and the $nL$ eigenstates of the potential $\Sigma_g^{+}$ [Eq.~\eqref{eq: potential}] obtained from the indicated sources.  Also computed are matrix elements indicating the spatial extent of the states.  Fine-structure effects have not yet been included.}
    \label{tab:sq-tetraquarks}
\end{table*}

\begin{table*}[]
\centering
    \renewcommand{\arraystretch}{1.5}
\begin{tabular}{ccccccc}
\hline\hline
BO state & Potential & $m_{\delta}$ (GeV) & $M_0$ (GeV) & $\langle 1/r\rangle^{-1}$ (fm) & $\langle r \rangle$ (fm) & $\text{Multiplicity}\times J^{PC}$ \\\hline
\multirow{2}{*}{$\Sigma_g^{+}(1S)$}
& LQCD~\cite{Morningstar:2019}  & 0.950    & 2.219 & 0.359  & 0.478 &          \\
%\cline{2-6}
& RQM~\cite{Ebert:2005xj}  & 1.203 & 2.183 & 0.245  & 0.345  & $0^{++},1^{+-},2^{++}$ \\\cline{1-6}
\multirow{2}{*}{$\Sigma_g^{+}(2S)$} 

& LQCD~\cite{Morningstar:2019}  & 0.950    & 2.909 & 0.547 & 0.881 &           \\
%\cline{2-6}
& RQM~\cite{Ebert:2005xj}  & 1.203 & 2.956 & 0.471  & 0.771  &  \\\hline%\hline
\multirow{2}{*}{$\Sigma_g^{+}(1P)$} 

& LQCD~\cite{Morningstar:2019}  & 0.950    & 2.656 & 0.627  & 0.723 &         \\
%\cline{2-6}
& RQM~\cite{Ebert:2005xj}  & 1.203   & 2.786 & 0.536 & 0.628  & $(0,1,2)^{-+},\ (2\times 1,2,3)^{--}$ \\\cline{1-6}
\multirow{2}{*}{$\Sigma_g^{+}(2P)$} 

& LQCD~\cite{Morningstar:2019}  & 0.950    & 3.233 & 0.793  & 1.068 &         \\
%\cline{2-6}
& RQM~\cite{Ebert:2005xj}  & 1.203   & 3.316 & 0.718 & 0.983  &  \\\hline%\hline
\multirow{2}{*}{$\Sigma_g^{+}(1D)$}

& LQCD~\cite{Morningstar:2019}  & 0.950    & 3.004 & 0.843  & 0.928 & $(0,1,2\times 2,3,4)^{++}, (1,2,3)^{+-}$         \\
%\cline{2-6}
& RQM~\cite{Ebert:2005xj}  & 1.203  & 3.141 & 0.767  & 0.849 &  \\\hline\hline
\end{tabular}
 \caption{Masses $M_0$ and corresponding $J^{PC}$ eigenvalues of $s\bar s s\bar s$ dynamical-diquark states, computed using the given diquark masses $m_{\delta}$ and the $nL$ eigenstates of the potential $\Sigma_g^{+}$ [Eq.~\eqref{eq: potential}] obtained from the indicated sources.  Also computed are matrix elements indicating the spatial extent of the states.  Fine-structure effects have not yet been included.}
 \label{tab:ss-tetraquarks}
\end{table*}

\begin{table*}[]
\centering
\renewcommand{\arraystretch}{1.5}
\begin{tabular}{c@{  }l|@{ }c@{  }l}
\hline\hline
\multicolumn{1}{c}{$J^{PC}$ \ } & \multicolumn{1}{c}{$P$-Wave Mass} &
\multicolumn{1}{@{\!\!}|c}{\, $J^{PC}$ \ } & \multicolumn{1}{c}{$D$-Wave Mass} \\
\hline
$1^{--}$ & $M_0(1P)+\kappa_{ss} - 3 V_{LS} - \frac{28}{5} V_T$  
  &
$0^{++}$ & $M_0(1D)+\kappa_{ss} - 6 V_{LS} - 8 V_T$  \\ %\hline
$0^{-+}$ & $M_0(1P)+\kappa_{ss} - 2 V_{LS} - 8 V_T$ 
   &
$1^{++}$ & $M_0(1D)+\kappa_{ss} - 5 V_{LS} - 4 V_T$   \\%\hline
$1^{-+}$ & $M_0(1P)+\kappa_{ss} - V_{LS} + 4 V_T$
   &
$1^{+-}$ & $M_0(1D)+\kappa_{ss} - 3 V_{LS} - 4 V_T$  \\%\hline
$2^{--}$ & $M_0(1P)+\kappa_{ss} - V_{LS} + \frac{28}{5} V_T$ \hspace{0.5em} &
$2^{++}$ & $M_0(1D)+\kappa_{ss} - 3 V_{LS} + \frac{12}{7} V_T$ 
\\%\hline
$1^{--}$& $M_0(1P)+\kappa_{ss}$
 &
 $2^{+-}$ & $M_0(1D)+\kappa_{ss} - V_{LS} + 4 V_T$  
  \\%\hline
$2^{-+}$ & $M_0(1P)+\kappa_{ss} + V_{LS} - \frac{4}{5} V_T$  &
$2^{++}$ & $M_0(1D)+\kappa_{ss}$ 

  \\%\hline
$3^{--}$ & $M_0(1P)+\kappa_{ss} + 2 V_{LS} - \frac{8}{5} V_T$  & 
$3^{++}$ & $M_0(1D)+\kappa_{ss} + 0 V_{LS} + \frac{32}{7} V_T$    \\%\hline
    &   &
    $3^{+-}$ & $M_0(1D)+\kappa_{ss} + 2 V_{LS} - \frac{8}{7} V_T$
    
 \\%\hline
  &  &   
$4^{++}$ & $M_0(1D)+\kappa_{ss} + 4 V_{LS} - \frac{16}{7} V_T$   \\
\hline\hline
\end{tabular}
\caption{Masses of $s\bar s s\bar s$ tetraquark states in $P$-wave and $D$-wave configurations including spin-spin ($\kappa_{ss}$), spin-orbit ($V_{LS}$), and tensor ($V_T$) interactions defined in Eq.~(\ref{eq:hamiltonian}).  In addition, the two $J^{PC} = 1^{--}$ $P$-wave states have a mixing mass term of size $+\frac{8}{\sqrt{5}} V_T$, and the two $J^{PC} = 2^{++}$ $D$-wave states have a mixing mass term of size $+\frac{8}{\sqrt{7}} V_T$.}
\label{tab:mass-splittings}
\end{table*}

We also present results for expectation values relevant to the size of the state in Tables~\ref{tab:sq-tetraquarks}-\ref{tab:ss-tetraquarks}.  While \(\langle r \rangle\) provides an estimate of the overall spatial extent of the state, \(\langle 1/r \rangle^{-1}\) probes the short-distance structure of the wave function and is particularly informative in the presence of a Cornell-type potential.  In this framework, the Coulombic term \(-\alpha/r\) dominates the short-distance interaction, while the linear term \(\sigma r\) governs long-range confinement.  Thus, \(\langle 1/r \rangle^{-1}\) effectively measures the average strength of the short-range attraction, while \(\langle r \rangle\) captures the extent of the confining color flux tube.   Such distinctions are useful for differentiating compact $\de\bde$ states from hadronic molecules, which generally exhibit both a larger \(\langle r \rangle\) and a larger \(\langle 1/r \rangle^{-1}\) (typically \(\gtrsim 1~\mathrm{fm}\)).

To interpret our results, we compare the predicted tetraquark masses with available experimental candidates and examine their compatibility with observed resonances.  Additionally, we contrast our findings with those obtained from alternative theoretical approaches, such as LQCD, QCD sum rules, and constituent-quark models, in order to assess the robustness and distinctive features of the dynamical-diquark framework.

\paragraph{$1S$ $s\bar s q\bar q$ Tetraquarks.} 
The mass spectrum and quantum numbers presented in Table~\ref{tab:sq-tetraquarks} suggest that the observed $2^{++}$ mesons $f_2(1910)$ and $f_2(1950)$ are promising candidates for $\Sigma^+_g(1S)$ $s\bar s q\bar q$ tetraquark states.  This assignment gains further plausibility when considering the difficulty of placing these states within conventional $q\bar{q}$ meson nonets (Sec.~63 of Ref.~\cite{ParticleDataGroup:2024cfk}).  In particular, the large $f_2(1950)$ decay width (464~MeV) and its observation in the $K \bar{K}$  decay channel~\cite{Belle:2003xlt} suggest that this meson possesses exotic quark content, with a significant $s\bar s$ component.

It is worth mentioning that the $\Sigma^+_g(1S)$ mass prediction using the LQCD potential of Ref.~\cite{Morningstar:2019} aligns with that of $X(1835)$ (which is also close to the $p\bar{p}$ threshold).  Nevertheless, its quantum numbers $0^{-+}$~\cite{BESIII:2015xco} (in particular, $P \! = \! -1$) disfavor its interpretation as an $S$-wave $s\bar sq\bar q$ tetraquark within the dynamical diquark model.

\paragraph{$1S$ $s\bar s s\bar s$ Tetraquarks.}
The predicted mass eigenvalue and allowed $J^{PC}$ values in Table~\ref{tab:ss-tetraquarks} suggest that one of the observed $2^{++}$ mesons $f_2(2300)$ and $f_2(2340)$ is a $s\bar s s\bar s$ $1S$ tetraquark.  It is worth noting that the latter state is also predicted to be an $s\bar s s\bar s$ tetraquark in Ref.~\cite{Liu:2020lpw}.

\paragraph{$1P$ $s\bar s q\bar q$ Tetraquarks.} The mass eigenvalue for $\Sigma^+_g(1P)$ listed in Table \ref{tab:sq-tetraquarks} is close to that of $X(2370)$ as well as to $X(2500)$ and $X(2600)$ (depending upon the particular potential used), although the quantum numbers $J^{PC}$ require experimental confirmation.  A recent potential-model analysis~\cite{Lodha:2024yfn} shows good agreement with our results, predicting the lightest $P$-wave states \( 0^{-+} \) and \( 1^{--} \) to lie at 2.3~GeV and 2.6~GeV, respectively. According to a recent QCD sum-rule analysis~\cite{Wang:2025nme}, $X(2370)$ is also interpreted as a $0^{-+}$ $P$-wave $s\bar s q\bar q$ tetraquark.  However, Ref.~\cite{Liu:2020lpw} interprets $X(2500)$ as a $0^{-+}$ $s\bar s s\bar s$ $P$-wave~ state, while  $X(2600)$ is described as a $2^{-+}$ $s\bar s q\bar q$ tetraquark in Ref.~\cite{Wang:2025nme}, and as a glueball in Refs.~\cite{Zhang:2022obn,Giacosa:2023fdz}.

We also comment upon the well-established resonance $\phi(2170)$, which is interpreted as an $s\bar s q\bar q$ tetraquark within QCD sum rules~\cite{Agaev:2019coa}.  An $s\bar s s\bar s$ $P$-wave tetraquark configuration for $\phi(2170)$ is not supported within the framework of the nonrelativistic quark model~\cite{Liu:2020lpw}.  On the other hand, an analysis of its decay channels supports its interpretation as an $s\bar s s\bar s$ tetraquark~\cite{Ke:2018evd}.  Although our model predictions estimate the mass of $s\bar s q\bar q$ $1P$ tetraquark state to be approximately 150-350~MeV higher in mass than $\phi(2170)$, it remains plausible to interpret $\phi(2170)$ as an $s\bar s q\bar q$ tetraquark when fine-structure spin-dependent effects are considered for the \( J^{PC} = 1^{--} \) state, assuming that such effects are similar in size to those observed in the hidden-charm sector~\cite{Giron:2020fvd,Giron:2020wpx}.

\paragraph{$1P$ $s\bar s s\bar s$ Tetraquarks.} 
$s\bar s s\bar s$ tetraquark states have been systematically explored in various theoretical frameworks aimed at understanding their internal structure and possible experimental signatures.  QCD sum-rule analyses have been employed to determine the masses and quantum numbers of these exotic configurations~\cite{Su:2022eun}, with specific attention to the $1^{--}$ sector in Ref.~\cite{Xin:2022qnv} and the $2^{++}$ sector in Ref.~\cite{Dong:2023}.  Meanwhile, these states have been studied in the nonrelativistic quark model, both using the diquark approximation (including numerical calculations of the decay properties)~\cite{Lodha:2024qby} and not using this approximation~\cite{Liu:2020lpw}.

We present a comparison in Fig.~\ref{fig:ss} between the dynamical diquark model (without fine-structure spin interactions) and the predictions of diquark-based QCD sum-rule calculations~\cite{Su:2022eun} (see also Refs.~\cite{Wang:2019nln,Xin:2022qnv}) and a diquark-based potential model~\cite{Lodha:2024qby}.  We observe that our results for $0^{++}$ and $1^{--}$ (again, neglecting spin-dependent effects) using the LQCD potential~\cite{Morningstar:2019} are in good agreement with Ref.~\cite{Lodha:2024qby}.  The differences in mass predictions for tetraquarks between QCD sum rules and potential models can be attributed to various underlying assumptions and their sensitivity to key parameters.  These parameters in QCD sum rules include the mass of the strange quark, the Borel window, and the continuum threshold \( s_0 \).  Interestingly, an older QCD sum-rule prediction~\cite{Chen:2008ej} for the lowest $1^{--}$ $s\bar s s\bar s$ tetraquark mass, $2.3\pm 0.4$ GeV, also aligns well with our prediction within their large theoretical uncertainties.

As mentioned above, no current experimental evidence fixes the values of the spin-dependent fine-structure operator coefficients in the $s\bar s s\bar s$ sector.  Here, purely for illustration purposes, we adopt hints from the hidden-charm $P$-wave sector~\cite{Giron:2020fvd} that $V_{LS} > V_T > 0$, and order the states according to the assumption $V_{LS}\gg V_{T}>0$.  The mass expressions for the $P$-wave $s\bar s s\bar s$ states are collected in Table~\ref{tab:mass-splittings}.  The \(1^{--}\) state based upon the state $X_2$ in Eq.~(\ref{eq:Swavediquark}) undergoes the largest mass shift due to the large \(-3V_{LS}\) term in the mass formula, making it one of the lightest among the $P$-wave states (although note that it mixes through $V_T$ with the $1^{--}$ state based upon $X^\prime_0$), followed by the $0^{-+}$ state.  The \(3^{--}\) and \(2^{-+}\) states, which have positive $V_{LS}$ coefficients, are expected to be the heaviest.  In the limit $V_{LS} \gg V_T$ the states \(1^{-+}\) and \(2^{--}\) become degenerate, although this degeneracy is lifted upon the inclusion of the tensor interaction.

A comparison of our mass expressions for the $P$-wave $s\bar s s\bar s$ states presented in Table~\ref{tab:mass-splittings} with those derived from the potential model and from QCD sum rules ($\lambda$ modes only, {\it i.e.}, those with no orbital excitation within $\de$ or $\bde$) reveals a consistent pattern: In both cases, the lightest $P$-wave state carries the quantum numbers $0^{-+}$, while the heaviest is $3^{--}$ (see Fig.~\ref{fig:ss}).  Notably, a similar hierarchy with $M_{0^{-+}} < M_{1^{--}}$ also emerges in our model if the condition $V_{LS} < \frac{12}{5} V_T$ is satisfied.

While we do not advocate here for particular values of the fine-structure parameters, we offer a few values taken from the literature to indicate their expected sizes.  One may estimate the parameter $\kappa_{ss} \approx 23$ MeV by utilizing the mass splittings observed in strange baryons (see Refs.~\cite{Karliner:2014gca,Maiani:2022psl}).    On the other hand, the remaining spin-dependent parameters, $V_{LS} \approx 29\pm 5$ MeV and $V_T \approx -1 \pm 2 $ MeV,  can be obtained from QCD sum-rule results for the tetraquark mass spectrum given in Ref.~\cite{Su:2022eun}.   The corresponding values obtained in a charm-sector fit~\cite{Giron:2020fvd} are $43$-$45~{\rm MeV}$, $43$-$49~{\rm MeV}$ and $3$-$ 6~{\rm MeV}$, respectively.

In the $D$-wave sector, Table~\ref{tab:mass-splittings} and the assumption $V_{LS}\gg V_{T}>0$ indicate that the two lightest states are $0^{++}$ and $1^{++}$, the two heaviest are $4^{++}$ and $3^{+-}$, and two degenerate pairs occur: $1^{+-}$ and $2^{++}$ [the latter based upon $X_2$ in Eq.~(\ref{eq:Swavediquark})], and $2^{++}$ (based upon $X^\prime_0$) and $3^{++}$.

\section{Conclusions}\label{sec:Concl}

The dynamical diquark model treats multiquark had\-rons as being composed of color-triplet diquark (and triquark) quasiparticle subcomponents that can briefly achieve a static configuration describable in terms of the Born-Oppenheimer approximation.  Key in the plausibility of the model is that the diquarks are substantially heavier than the mass equivalent of the light degrees of freedom in the color flux tube connecting them.

Here we have pushed the model to the limit of its feasibility by treating diquarks containing strange quarks as heavy, in order to explore what vestiges of the model persist in the $1.9$-$3.2$~GeV region.  In this mass range, several meson resonances with unusual properties have been observed, such as $\phi(2170)$, $f_2(2340)$, and $X(2370)$, and are often interpreted as candidates for tetraquarks, hybrids, or glueballs.

The dynamical diquark model, by limiting inter-di\-quark color forces to occur in the QCD color triplet, predicts a very specific ordering of radial and orbital ($nL$) excitations when one chooses any definite potential-energy function connecting them (in this work, taken from lattice-QCD simulations and from quark models).  We have also extracted numerical values for the masses of $(sq)$ and $(ss)$ diquarks from such previous works, and used these potentials and masses to solve the resulting Schrödinger equations.  We find, for example, that $f_2(1950)$ is a good $1S$ $s\bar s q\bar q$ candidate, either of $f_2(2300)$ or $f_2(2340)$ fits well as a $1S$ $s\bar s s\bar s$ candidate, states such as $X(2500)$ are plausible $1P$ $s\bar s q\bar q$ candidates, and $\phi (2170)$ is more likely a $1P$ $s\bar s q\bar q$ than a $1P$ $s\bar s s\bar s$ state.

The limitation to (antisymmetric) color-triplet diquarks also greatly restricts the specific spectra of $J^{PC}$ quantum numbers in each $nL$ multiplet, and this specificity provides strong constraints, particularly in the $s\bar s s\bar s$ sector due to the Pauli exclusion principle.  Here, the model predicts the existence of precisely 3 approximately degenerate states in $nS$ multiplets (including the ground-state $1S$ multiplet), with $J^{PC} = 0^{++}, 1^{+-}$, and $2^{++}$.  Under mild assumptions on the nature of the spin-dependent couplings, we also predict the mass ordering of states within $nP$ and $nD$ $s\bar s s\bar s$ multiplets.

This work presents the results of a deliberately simplistic model.  In particular, we have ignored effects due to the coupling of $s\bar s q\bar q$ and $s\bar s s\bar s$ resonances to di-meson thresholds, including di-meson pairs ({\it e.g.}, $K^{(*)} \bar K^{(*)}$ or $\phi \,\eta^\prime$) to which the tetraquark states may decay.

Such modifications may be incorporated rigorously, generalizing the adiabatic nature of the Born-Oppenheimer approximation to produce the so-called {\em diabatic\/} dynamical diquark model~\cite{Lebed:2022vks}.  This generalization is ideally suited to studying decay widths~\cite{Lebed:2024rsi} and resonant line shapes~\cite{Lebed:2023kbm} of such states, and constitutes a major thrust for future work.

These effects can be quite dramatic.  As shown in the $c\bar c$ sector~\cite{Lebed:2022vks}, the exotic state $X(3872)$ has a dominant $D\bar{D}^*$ component, comprising approximately 91\% of its structure,  but still has a non-negligible $\de\bde$ contribution even though the di-meson threshold is only about 0.04 MeV higher than the $X(3872)$ mass.  Including the di-meson coupled channels can actually reduce the mass eigenvalue by tens of MeV\@.  The same analyses show that, when the $S$-wave $\de\bde$  component dominates, the state exhibits a preferential coupling to $S$-wave di-meson channels over $D$-wave channels, even when the $D$-wave threshold lies closer to the mass eigenvalue.  A similar situation is expected in the $S$-wave $s\bar s$ tetraquark sector.
    
In the $c\bar c$ sector, a diabatic coupled-channel approach is also used to compute mass shifts and decay widths resulting from open di-meson thresholds~\cite{Lebed:2024rsi}.  Notably, the mass corrections for $J^{PC}=2^{++}$ states remain below 10~MeV and the widths are typically sub-MeV, indicating minimal impact from continuum coupling.  In contrast, $1^{--}$ states exhibit more substantial continuum dressing: Their widths can reach tens of MeV, as seen for hidden-charm vectors ($Y$) decaying into open-charm  meson pairs.  These quantitative findings illustrate that while tensor states are relatively robust against continuum effects, vector states [like $\phi(2170)$] may be significantly modified.  For instance, Ref.~\cite{Coito:2009na} employs a multichannel $T$-matrix formalism to demonstrate that the $\phi(2170)$ resonance can be dynamically generated through strong coupling between a confined $s\bar{s}$ seed state and open di-meson channels such as $K^*\bar{K}_1(1270)$.  This interaction produces a resonance near the observed mass despite the substantial $K_1$ decay width, and the diabatic model likely produces a similar result.

We also focused exclusively upon tetraquarks in this paper.  However, as noted above, the dynamical diquark model can also be used to produce pentaquark states as diquark-triquark composites.  An application of this approach to the hidden-strangeness baryon sector can address phenomena such as the cusplike structure seen near 1900~MeV in $\gamma \, p \to K^+ \Sigma^0$ by the BGOOD Experiment~\cite{Jude:2020byj}.

Meanwhile, a detailed calculation of the fine-structure corrections required to resolve the multiplets studied here requires either robust predictions from lattice QCD, quark models, or QCD sum rules, or additional hints from experiment.  With regard to the latter, all measurements of such states, whether they eventually turn out to be hidden-strangeness tetraquarks, hybrids, or glueballs (or even peculiar highly excited conventional mesons) are of great interest to ongoing experimental efforts at BESIII and JLab, and in future experiments at the EIC.

\begin{acknowledgments}
RFL acknowledges support by the National Science Foundation (NSF) under Grants No.\ PHY-2110278 and PHY-2405262.  SJ acknowledges support by the U.S.\ Department of Energy ExoHad Topical Collaboration under Grant No.\ DE-SC0023598.  This work contributes to the goals of the ExoHad Collaboration.
\end{acknowledgments}

%\clearpage

\bibliographystyle{apsrev4-2}
\bibliography{ssss}
\end{document}